\documentclass[graybox]{svmult}

\usepackage{mathptmx}       
\usepackage{helvet}         
\usepackage{courier}        
\usepackage{type1cm}        
\usepackage{makeidx}         
\usepackage{graphicx}        
\usepackage{multicol}        
\usepackage[bottom]{footmisc}

\makeindex             

\begin{document}

\title*{Kinetic Exchange Models in Economics and Sociology}
\author{Sanchari Goswami and Anirban Chakraborti }
\institute{Sanchari Goswami \at S.N. Bose National Centre for Basic Sciences, JD Block, Sector III, Salt Lake City,
Kolkata 700098, INDIA \email{sanchari.goswami@bose.res.in}
\and Anirban Chakraborti \at School of Computational and Integrative Sciences, Jawaharlal Nehru University, New Delhi-110067, INDIA
\email{anirban@mail.jnu.ac.in}}
\maketitle

\abstract{In this article, we briefly review the different aspects and applications of kinetic exchange models in economics and sociology. Our main aim is to show in what manner the kinetic exchange models
for closed economic systems were inspired by the kinetic theory of gas molecules. The simple yet powerful framework of kinetic theory, first proposed in 1738, led to the successful development of statistical physics of gases towards the end of the 19th century. This framework was successfully adapted to modeling of wealth distributions in the early 2000's. In later times, it was 
applied to other areas like firm dynamics and opinion formation in the society, as well. We have tried to present the flavour of the several models proposed and their applications, intentionally leaving out the intricate mathematical and technical details.
}

\section{Introduction}
\label{sec:1}
The aim of statistical physics is to study the physical properties of macroscopic systems consisting of a large 
number of particles. In such large systems, the number of particles is of the order of Avogadro number. Thus it 
is extremely difficult to have a complete microscopic description of such a system, both experimentally and 
by the way of solving equations of motion. In spite of the complexity of such systems, they exhibit some macroscopic 
observable quantities, which represent averages over microscopic properties \cite{Mandl,terHaar1995,Sethna_A}. 

A society can be described as a group of people sharing the same 
geographical or social territory and involved with each other by means of sharing different aspects of life. 
In sociology, a branch of social sciences, one studies the human social behavior in a society. 
Economics is another branch of the social sciences which analyzes the production, distribution, and consumption of 
goods and services.
Since the society is usually formed with a very large number of people, the study of an individual is extremely difficult. 
However in various cases, one can observe and characterize some average behaviour of the people, e.g., in case of a voting a large number of people selects a particular opinion. Similar to many physical phenomena, quite well-understood by physicists, it has been found that a study of crime, a social phenomenon, displays a first-order transition between states of high and low crime rates as a function of severity 
of the criminal justice system. Also, a model of marriage, another social phenomenon, show critical behaviour such that the relation among marriage rates, economic incentives and social pressures show a surface similar to a $P$-$V$-$T$ surface of a fluid. Also, the dynamical nature of interaction of any economic sector which is composed of a large number of cooperatively
interacting agents, has many features in common with the interacting systems of statistical physics.
These na\"{\i}vely suggest that study of society as viewed  
by the economists and sociologists, can also be done using the tools of statistical mechanics developed by the physicists. The application of statistical mechanics to the fields of economics and sociology have resulted in the 
interdisciplinary fields 
namely ``econophysics''  \cite{SCCC} and 
``sociophysics'' \cite{sen_chak_book}.
According to P. Ball \cite{ball},
\begin{quotation}
At face value, there might seem to be little room left for statistical physics to make
a realistic contribution. But if there is one message that emerges clearly from this
discipline, it is that sometimes the details do not matter. That, in a nutshell, is what
is meant by universality. It does not matter that the Ising model is a ridiculously crude
description of a real fluid; they both have the same behaviour at the critical point
because in that circumstance only the broad-brush features of the system, such as the
dimensionality and range of particle interactions, determine the behaviour.
\end{quotation}

The kinetic exchange model is one of the simplest models in
statistical mechanics, which derives the average macroscopic behaviours
from the microscopic properties of particles. The kinetic exchange model
is in general based on the exchange of energy among particles due
to elastic collisions occuring among them. Bernoulli, in 1738, gave a complete description of the movement 
and activities of gas molecules in \textit{Hydrodynamica} which is well known as ``Kinetic theory of gases''. 
This attempt was later developed and formalized by several other pioneers of \textquoteleft Statistical Thermodynamics\textquoteright, such as Clausius, Maxwell, Boltzmann, Planck, and Gibbs. In this paper, we will present some existing models in several fields of, not only natural sciences but also social sciences, such as economics and sociology \cite{CUP}. 

\section{Kinetic Exchange models in Economics}
\label{sec:2}
An economy can be studied in various ways. For example, one can study the economy in the light of individual's
wealth as well as production of goods or wealth by firms in that economy. 
The economy consists of a large number of firms populated by workers. By
firms we mean production units, each and every one of which capable of producing any kind of
goods and services.

The famous Italian economist Vilfredo Pareto, in 1897, 
observed that the income distribution in Europe follow a power-law tail \cite{Pareto:1897}. 
The tail-end distribution of income is given as,
\begin{equation}
 p(m) \sim m^{-(1+\nu)},
\end{equation}
where $\nu$ is called the Pareto exponent. The value of the exponent as measured by Pareto for 
different kingdoms and countries varied between $1.1$ to $1.7$ \cite{Pareto:1897}. Pareto also observed that roughly $80$\% of the total wealth is limited to the hands of only $20$\% people of the society; this signifies that there is a small finite number of very very rich people in a society. 

Several surveys were done to verify Pareto law. Japanese, Australian and Italian personal income distribution have been shown to have a log-normal distribution for the lower income range and a power-law tail at higher income portion \cite{Souma, Matteo, Clementi}. 
In India, studies revealed that the income of rich people follow 
power-law distribution \cite{Sinha}. 
Similar thing is observed for the income and wealth distribution in modern USA and UK \cite{Silva, Dragu} and other countries.
All these studies show the evidence of the power law tail but 
the Pareto exponent is found to vary between $1$ and $3$ 
\cite{Souma, Matteo, Clementi, Sinha, Silva, Dragu, Mandelbrot:1960,EWD05,ESTP,SCCC,Yakovenko:RMP, datapap}.

\begin{svgraybox}
In any society or country, one finds that the total wealth remains fairly constant on a
longer time scale than its movement from individual
to individual. This is because the dynamics of the latter occurs at shorter time
scales (e.g. daily or weekly). This in turn results in very robust type of wealth distributions. Empirical data for society show a
small variation in the value of the power-law exponent at 
the \textquoteleft tail\textquoteright of the distribution, while it equals
to unity for firms. 
\end{svgraybox}

The interesting question is then, why is such ``universal'' behaviour as the widespread Pareto law, observed in the case of wealth distribution in the society. To this aim, a number of models have been proposed to reproduce these observed features, specifically to obtain
a power-law tail as was observed in empirical data. Many of
these models have been inspired by the kinetic theory of gas-like
exchanges. Notably, in 1960, the mathematician and economist Mandelbrot, wrote:
\begin{quotation}
There is a great temptation
to consider the exchanges of money which occur in economic interaction as analogous to the exchanges of energy
which occur in physical shocks between molecules. In the
loosest possible terms, both kinds of interactions should
lead to similar states of equilibrium. That is, one should
be able to explain the law of income distribution by a
model similar to that used in statistical thermodynamics:
many authors have done so explicitly, and all the others
of whom we know have done so implicitly.
\end{quotation}


\subsection{Ideal gas-like Kinetic Wealth Exchange Models (KWEM)}
\label{subsec:2}

\begin{figure}
\sidecaption[t]
\includegraphics[scale=.45]{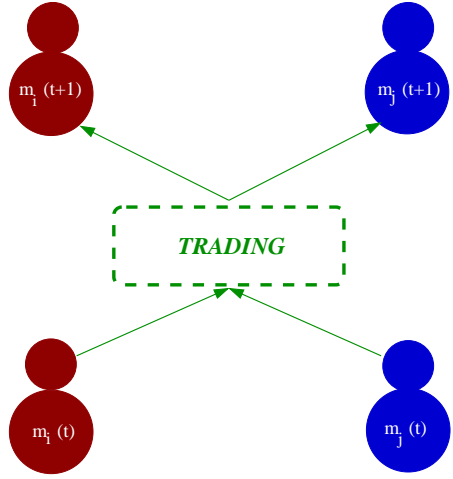}
\caption{A typical example of two agents $i$ and $j$ taking part in a trading process. 
Agent $i$ and $j$ have wealth $m_i(t)$ and $m_j(t)$ at time $t$. After a trading their wealth become $m_i(t+1)$ and 
$m_j(t+1)$ respectively.
}
\label{exchange}
\end{figure}

A trading process may be realized in a manner similar to the gas molecules exchanging energy in the kinetic theory of gases, where now
a pair of traders exchange wealth, respecting local conservation in any trading
\cite{marjitIspolatov,Dragulescu:2000,Chakraborti:2000,Chatterjee:rev,Chakrabarti:2010,Chatterjee:2010}. 
These models have a
microcanonical description and nobody ends up with negative
wealth (i.e., debt is not allowed). Thus, for two agents $i$ and
$j$ with wealth $m_{i}(t)$ and $m_{j}(t)$ at time $t$, the general trading
is given by:
\begin{equation}
\label{mdelm}
m_i(t+1) = m_i(t) + \Delta m; \  m_j(t+1) = m_j(t) - \Delta m;
\end{equation}
time $t$ changes by one unit after each trading. A typical wealth exchange process is shown in Fig. \ref{exchange}. 

\subsubsection{Model with no saving}
In a simple conservative model proposed by  Dr\u{a}gulescu and Yakovenko (DY model) 
\cite{Dragulescu:2000},  
$N$ agents 
exchange wealth randomly keeping the total wealth $M$ constant. 
The simplest model considers a random fraction of total
wealth to be shared:
\begin{equation}
 \Delta m = \epsilon_{ij} (m_{i}(t) + m_{j}(t)) - m_{i}(t),
\end{equation}
where $\epsilon_{ij}$ is a random fraction $(0 \leq \epsilon_{ij} \leq 1)$.
The steady-state ($t \rightarrow \infty$) wealth follows a Boltzmann-Gibbs distribution:
$P(m)=(1/T)\exp(-m/T)$; $T=M/N$, 
a result which is robust and independent of the topology of the (undirected)
exchange space 
\cite{Chatterjee:rev}.

\begin{svgraybox}
The Boltzmann-Gibbs distribution, a fundamental law of equilibrium statistical 
mechanics, states
that the probability $P(\epsilon)$ of finding a physical system or
subsystem in a state with the energy $\epsilon$ is given by the
exponential function
$$P(\epsilon) = ce^{−\epsilon/T}.$$ Here the conserved quantity is the total energy.
\end{svgraybox} 

If $m_{1} > m_{2}$ and the agents share some random fraction of 
wealth $2m_{2}$ and not of the total $(m_{1} + m_{2})$, which indicates trading at
the level of lower economic class in the trade, then all the wealth in the market drifts
to one agent drastically
\cite{Chakraborti2002, Hayes}. In \cite{lopez}, different approaches to obtain the exponential Boltzmann-Gibbs 
distribution have been addressed and a new operator in the framework of functional iteration theory has been 
proposed. It shows the exponential distribution to be ubiquitous in the 
framework of many multi-agent systems, not only economic ones but more diverse ones which have
some economic inspiration included.

\subsubsection{Model with uniform saving}
An additional concept of {\it saving propensity} was considered first by Chakraborti and 
Chakrabarti \cite{Chakraborti:2000} 
(CC model hereafter). Here, the agents save a fixed fraction $\lambda$ of their wealth
when  interacting  with another agent. Thus,
two agents with initial wealth $m_i(t)$ and $m_j(t)$ at time $t$ interact such that 
they end up with wealth $m_i(t+1)$ and $m_j(t+1)$
given by
\begin{eqnarray}
\label{fmij}
m_i(t+1)=\lambda m_i(t) + \epsilon_{ij} \left[(1-\lambda)(m_i(t) + m_j(t))\right], \nonumber \\
m_j(t+1)=\lambda m_j(t) + (1-\epsilon_{ij}) \left[(1-\lambda)(m_i(t) + m_j(t))\right];
\end{eqnarray}
$\epsilon_{ij}$ being a random fraction between $0$ and $1$, modeling the stochastic nature
of the trading.
It is easy to see that the $\lambda=0$ case is equivalent to the DY model -- the market is non-interacting 
in this case, and the most probable wealth per agent is $0$ here. The market is again non-interacting 
for $\lambda = 1$ when the most probable wealth per agent is $M/N$. We have a so-called 
\textquoteleft interacting \textquotedblright market when $\lambda$ has any non-vanishing value between $0$ and $1$. 
The steady state distribution $P(m)$
is exponentially decaying on both sides.
It is interesting to note that, the most probable value for such $\lambda$'s is something in between $0$ 
and $M/N$ so that the fraction of deprived people decrease with saving
fraction $\lambda$ and most people end up with some finite fraction of the average wealth in the market. 
This is a ``self-organizing'' feature of the market.
This results in completely different types of wealth 
distribution curves, very well approximated  by Gamma distributions~\cite{Patriarca:2004,Repetowicz:2005,Lallouache:2010}
given by, 
\begin{equation}
 P(m) = Cm^{\alpha}\exp(-m/T),
\end{equation}
where $T = \frac{1}{\alpha + 1}$ and $C = \frac{(\alpha +1)^{\alpha +1}}{\Gamma(\alpha +1)}$.
The exponent $\alpha$ is related to the saving propensity $\lambda$ by the relation :
\begin{equation}
 \alpha = \frac{3 \lambda}{1-\lambda}.
\end{equation}
The $\lambda=0$ limit can be verified from the above results.
This fits well to empirical data for low and middle wealth 
regime~\cite{Souma, Matteo, Clementi, Sinha, Silva, Dragu, datapap}.
The model features are somewhat similar to Angle's work~\cite{Angle}.
Obviously, the CC  model did not lead to the expected behaviour according to Pareto law.

In \cite{ChakrabartiABK,Chakraborti-prl,AJP}, the equivalence between kinetic wealth-exchange models 
and mechanical models of particles was shown and the universality of the underlying dynamics was studied both
through a variational approach based on the minimization of the Boltzmann entropy
and a microscopic analysis of the collision dynamics of molecules in a gas.
In case of systems with a homogeneous quadratic Hamiltonian and $N$ (effective) degrees of freedom, the
canonical equilibrium distribution is a gamma-distribution of order $N/2$.
For the CC model, the effective dimension $N = 2(1+\alpha) = 2\frac{1+2\lambda}{1-\lambda}$ 
and therefore, the corresponding distribution has the special property that it becomes a Dirac-$\delta$ or 
fair distribution
when $\lambda \rightarrow 1$ or $N(\lambda) \rightarrow \infty$. 
\subsubsection{Model with distributed savings}
In a later  model  proposed by Chatterjee et. al.~\cite{Chatterjee:2004} 
(CCM model hereafter) it was assumed that the saving propensity has a distribution
and this immediately led to a wealth distribution curve 
with a Pareto-like tail having $\nu =1$. Here, 
\begin{eqnarray}
\label{mij}
m_i(t+1)=\lambda_i m_i(t) + \epsilon_{ij} \left[(1-\lambda_i)m_i(t) + (1-\lambda_j)m_j(t)\right], \nonumber \\
m_j(t+1)=\lambda_j m_j(t) + (1-\epsilon_{ij}) \left[(1-\lambda_i)m_i(t) + (1-\lambda_j)m_j(t)\right];
\end{eqnarray}
which are different from the CC model equations as $\lambda$'s are now agent dependent.
The steady state wealth distribution gave rise to a power law tail with exponent $2$.
Various studies on the CCM model have been made 
soon after~\cite{Chatterjee:2005,Mohanty:2006,Kargupta,ecoanneal,Toscani,
Chatterjee:2009,ChakrabartiASBK,ChakrabartiASBK2}.

Manna et. al. \cite{manna} used a 
preferential selection rule using a pair of continuously tunable 
parameters upon traders with distributed saving
propensities and was able to mimic the trend of  enhanced rates
of trading of the rich. The wealth distribution was found to follow Pareto law. It might be mentioned that in a similar context of preferential selection rules in wealth exchange processes, Iglesias et al. \cite{iglesias} had considered much earlier a model for the economy, where the poorest in the society at any stage
takes the initiative to go for a trade (random wealth exchange) with anyone else. Interestingly, in the
steady state, one obtained a self-organized poverty line,
below which none could be found and above which, a
standard exponential decay of the distribution (Gibbs)
was obtained.

\subsubsection{Extended CCM model}
In the extended CCM model \cite{AGhosh,sinha_chat}, a trade takes place between two agents in such a way that the investments of both agents 
are the same. For two agents $i$ and $j$ having wealth $m_i$ and 
$m_j$ respectively, the ``effective'' saving propensities are $\lambda_i = \frac{m_i}{m_i+m_j}$ and 
$\lambda_j = \frac{m_j}{m_i+m_j}$ respectively, which are functions of time. It is observed that in steady 
state, the wealth condenses to a single agent, a feature very similar to the results obtained by Chakraborti
\cite{Chakraborti2002}. By introducing taxation in the system not only condensation can be avoided but at the same time 
the model tends towards reality.
The tax is applied for the agents who have wealth greater than the average
wealth and this tax is collected periodically after a constant time interval. The total collected tax is 
then re-distributed over all the agents.
It is found that the distribution of wealth again has a power law tail with exponent $1.5$.

\subsection{Model with Phase Transition}
In \cite{threshold}, the authors introduced the concept of ``poverty line'', i.e., a 
threshold $\theta$, in the CCM model. A trade between two agents occurs as it is in the CCM model but with the restriction that at least one of the two agents
should possess wealth less than $\theta$. However, if all agents accumulate
wealth greater than $\theta$, then in such a situation the dynamics stops.
To continue the dynamics a perturbation is applied such that a
particle having energy above $\theta$ is selected randomly and its energy fully transfered
to any other particle. 
The maximum limit of the threshold value $\theta$ below which the dynamics is stopped within some
finite time, is the critical value of the threshold $\theta_c$.
The order parameter $O$ is defined as the 
average total number of agents having wealth less than $\theta$ i.e., $O = \int_{0}^{\theta} P(m)dm$, where
$P(m)$ is the wealth distribution. 
After a certain \textquoteleft relaxation time\textquoteright $\tau$, the system attains a steady state 
and several quantities are measured. If the order parameter $O$ is plotted against $\theta$, 
it is observed that after the point $\theta = \theta_c = 0.6075$
the order parameter increases.  The model thus has a ``phase transition'' near $\theta_c$ below which the
number of particles in the steady state goes to zero. Near the critical
point, the order parameter obeys a scaling form as $O \sim (\theta - \theta_c)^{\beta}$, where $\beta = 0.97$ 
is the order parameter exponent. Time variation of the 
order parameter has the scaling form $O(t) \sim t^{\delta}$ with exponent $\delta = 0.93$. Also 
a clear time scale divergence behavior is observed with scaling form $\tau \sim |\theta - \theta_c|^{-z}$.

\subsection{Nature of transactions in gas-like models with distributed savings}
\begin{figure}
\centering
\includegraphics[scale=0.45]{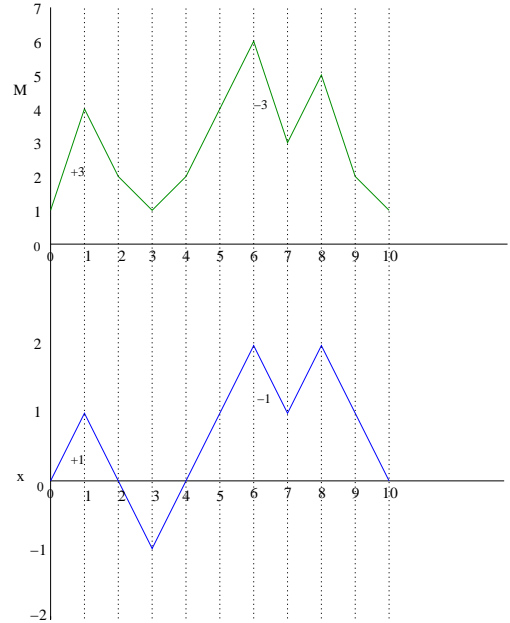}
\caption{Above : Plot of wealth $M$ of an agent in different steps. Below : Plot of the distance traveled $x$ 
in the gain-loss space by the corresponding walker. Note that, whatever be the amount of gain or loss, the step length of the walker is 
only $1$.
}
\label{step_money}
\end{figure}

The agent dynamics for models
with saving propensity can be studied  with emphasis on the nature of transactions, i.e., whether it
is a gain or a loss~\cite{Chatt-sen:2010}. 
In order to study the dynamics  of the transactions (i.e., gain or loss), a walk 
was conceived for the agents in an abstract one dimensional gain-loss space (GLS) where the agents conventionally
take a step towards right if a gain is made and left otherwise. Here the amount of gain or loss was not considered, 
i.e., whatever be the amount of gain or loss, the step length is only $1$. If it is a gain, the corresponding 
walker moves one step to the right and if it is a loss, walker moves one step to the left. For better understanding 
this is shown in Fig. \ref{step_money}.
It can be observed that in the CCM model, 
 the amount of wealth gained or lost by a tagged agent in a single interaction follows a 
 distribution  which is not symmetric in general,
well after equilibrium has been reached. 
The distribution depends strongly on the saving propensity of the agent. 
For example, an agent with larger $\lambda$ suffers more losses 
of less denomination compared to an agent with smaller $\lambda$, 
 although, in this case, the total wealth of the two agents has reached equilibrium, that is, each agent's wealth fluctuates around  a 
$\lambda$ dependent value. 

For such a walk, it can be found that $\langle x\rangle$, the distance traveled, scales linearly with 
time $t$ suggesting a ballistic 
nature of the walk for the CCM walk. Moreover, the slope of the  
$\langle x\rangle$ versus $t$ curves is dependent on $\lambda$; it is positive 
for small $\lambda$ and continuously goes to negative  values for larger values of $\lambda$. The slope becomes zero at
a value of $\lambda^* \simeq 0.469$.  In general for the CCM walk $\langle x^2 \rangle$ scales with $t^2$ .
For the CC model on the other hand, $\langle x^2 \rangle$ scaled with $t$ as in a random walk 
while $\langle x \rangle  \approx 0$.
The above results na\"{i}vely suggests that the walk in the GLS is like a biased random walk (BRW) 
(except perhaps at $\lambda^*$) for the CCM model while it is like a random walk (RW) for the CC model.

\subsection{Antipersistence Effect in CC/CCM Walk}
In \cite{goswami}, the exact nature of the walk associated with CC and CCM model was explored and it was 
shown through the effective bias $p$ associated with the walks, distribution of walk lengths at a stretch etc., 
that CCM is not a simple BRW and CC is 
not a simple RW. 

For BRW, the probability of direction reversal is simply $2p(1-p)$ which has a maximum value of $1/2$ 
for $p=1/2$. But for CCM, the direction reversal probability $f$ is greater than $1/2$ for all $\lambda < 1$ and 
$f \rightarrow 1/2$ for $\lambda \rightarrow 1$. Through further analysis of time correlation and other relevant 
quantities it was shown that direction reversal is preferred in these cases \cite{goswami}. 
In the equivalent picture of the walk in the abstract
space for gains and losses, it is similar to the fact that here individuals has a tendency to make a
gain immediately after a loss and vice versa. This so called ``antipersistence effect'' is in fact
compatible with human psychology where one can afford to incur a loss after a gain and will try to have a
gain after suffering a loss.

It was also shown in \cite{goswami} that the ``antipersistence effect'' is maximum for no saving and decreases
with saving. This is perhaps in tune with the human feeling of security associated with the
saving factor. In the CCM model, the saving propensity is randomly distributed and the
antipersistence effect occurs with a simultaneous bias that too depends on $\lambda$. 


\subsection{Firm Dynamics}
Size of a firm is measured by the strength of its workers. A firm grows when worker leaves another firm 
and joins it. The rate at which a firm gains or loses workers is called the ``turnover rate'' in economics 
literature. Thus there is a redistribution of workers and the corresponding dynamics can be studied. 
In the models of firm dynamics, one assumes the following facts : 
\begin{enumerate}
\item Any formal unemployment is avoided in the model. Thus one does not have to keep track of
the mass of workers who are moving in and out of the employed workers’ pool. 
\item The workers are treated as a continuous variable.
\item The definition that size of a firm is just the mass of workers working in the firm, is adopted.
\end{enumerate}

In firm dynamics models, we may make an analogy with the previous sub-sections that firms are agents and the number of workers in the firm is its wealth. Assuming no migration, 
birth and death of workers, the economy thus remains conserved. As the ``turnover rate'' dictates both the 
inflow and outflow of workers, we need another parameter to describe only the outflow. That parameter may 
be termed as ``retention rate'', which describes the fraction of workers who decide to stay back 
in their firm. This is identical to saving propensity in wealth exchange models, discussed earlier.

\subsubsection{Model with Constant Retention Rate}
In this model \cite{chak_int2}, the economy was considered to have $N$ firms and any firm could absorb any number of workers. Intially 
all firms have one unit of workers. The retention rate is denoted by $\lambda$. For this model, the 
retention rate of all firms are taken to be identical, as was in \cite{Chakraborti:2000}, which in reality is not true. The 
size of the $i$th firm $w_i$ ($i \leq N$). At each time, it was considered that $(1-\lambda)$ fraction of the workforce 
of $n$ firms (not $N$!), wanted to leave voluntarily or the firms wanted them to leave. The dynamics for the $i$th firm can be given as follows :
\begin{equation}
 w_i(t+1)=\lambda w_i(t) + \epsilon_{i(t+1)}(1-\lambda)\sum_j^n w_j(t),
 \label{cons_rr}
\end{equation}
where $\epsilon_{i(t+1)}$ are random variables which describes the fraction of workers actually moved to 
the $i$th firm at time $t+1$ among those who wanted to move. Note that, 
we use $t$ within the first bracket when referring to the endogenous variables\footnote{A classification 
of a variable generated by a statistical model that is explained 
by the relationships between functions within the model.} like the size of the firm $w_i(t)$
and the same in subscript when referring to the exogenous random variables\footnote{A variable whose 
value is determined outside the model in which it is used.} $\epsilon_{i(t)}$.

\runinhead{Restrictions on $\epsilon$}
\begin{enumerate}
\item $\sum_j^n \epsilon_j(t) = 1$ for all $t$ as the economy should be conserved.
\item Expectation $E(\epsilon_i)=1/n$ for all $i$ indicating that distributions of all $\epsilon_i$'s are identical.
\item If $n=2$, $\epsilon_i \sim [0,1]$ so that at the lower limit of $n$, CC/CCM can be got back.
\end{enumerate}

An exact solution was given in \cite{chak_int2} where it was assumed that all firms interact at every step. The 
steady-state distribution of the firms was shown to be
\begin{equation}
 f(w) = \lim_{\bar k \rightarrow \infty} \sum_{i=1}^{\bar k} \phi_i \exp(-\phi_i w) \prod_{i=1,j \neq i}^{\bar k} (\frac{\phi_j}{\phi_j-\phi_i}),
\end{equation}
where $\phi_i = \frac{1}{\lambda^i (1-\lambda)}$.
\subsubsection{Model with Distributed Retention Rate}
Here instead of a fixed retention rate, we consider distributed $\lambda$, i.e., Eq. \ref{cons_rr} can now be wriiten as
\begin{equation}
  w_i(t+1)=\lambda_i w_i(t) + \epsilon_{i(t+1)}\sum_j^n (1-\lambda_j) w_j(t).
  \label{dist_rr}
\end{equation}
The distribution of firm sizes can be  shown to be a power-law, by calculations similar to the one followed in \cite{pkm}.
\subsubsection{Model with Time-varying Retention Rate}
In this model, the retention rate $\lambda$ was taken to be a function of the evolving variable, the work-force $w$ \cite{chak_int2}. 
Thus Eq. \ref{cons_rr} can be modified in the following way,
\begin{equation}
 w_i(t+1)=\lambda(w_i(t)) w_i(t) + \epsilon_{i(t+1)}(1-\lambda(w_i(t)))\sum_j^n w_j(t).
\end{equation}
Following \cite{chak_int2} the functional form of $\lambda$ can be assumed as,
\begin{equation}
 \lambda(w) = c_1 (1-\exp(-c_2w));~~~~~~c_1,~c_2~~{\rm{are~constants}},
\end{equation}
which signifies a more realistic scenario that retention rate increases as current work-force increases.
This model leads to prominent bimodality in the
size distribution of firms \cite{chak_int2}. This has been empirically found
in the developing economies.

\section{Kinetic Exchange Models in Sociology}
Social systems 
offer some of the richest complex dynamical systems, which can be studied using the
standard tools of statistical physics. 
The study of Sociophysics became popular in the last part of 20th century 
\cite{AJP,Castellano, Galam, Stauffer}. 

Auguste Comte used the term \textquoteleft social physics\textquoteright in his 1842 work. He defined
social physics as the study of the laws of society or the science of civilization.
In particular,
Comte (1856) stated that,
\begin{quotation}
Now that the human mind has grasped celestial and 
terrestrial physics, mechanical
and chemical, organic physics, both vegetable and animal, there remains
one science, to fill up the series of sciences or observation -- social physics. This is
what men have now most need of...
\end{quotation}

\begin{svgraybox}
Emergence of consensus is an important
issue in sociophysics problems. Here, people interact to select an option among 
different options of a subject which may be vote, language, 
culture, opinion etc. This then leads to a state of consensus. In opinion formation, consensus is 
an ``ordered Phase'', where the most of the people 
have a particular opinion. Several models can be proposed to mimic the dynamics of opinion spreading. 
In the models of opinion dynamics, opinions are usually modeled as discrete or continuous variables 
and are subject to either spontaneous changes or changes due to binary interactions, 
global feedback and external factors (see \cite{Castellano} for a general review). 
\end{svgraybox}

However, in this paper, only kinetic exchange models of opinion dynamics, analogous to the ones in economics, 
will be discussed. These models are named after Lallouache, Chakrabarti, Chakraborti and Chakrabarti and are called  
LCCC model hereafter.
The opinions of individuals are assumed to be continuous variables in $[-1, 1]$ and
change due to binary interactions. Lallouache The tuning parameter in these models
is \textquoteleft conviction\textquoteright $\lambda$, which is 
similar to the \textquoteleft saving propensity\textquoteright as in KWEM. It determines
the extent to which one remains biased to its own opinion, while
interacting with the other. Unlike KWEM, there is no step-wise opinion conservation. 
\subsection{LCCC model}
In this model \cite{Lalla1, Lalla2}, opinion can be shared only in the two-body interaction mode. At any time $t$ a person $i$ is 
assigned with an opinion value $o_i(t) \in [-1,1]$. For two persons $i$ and $j$, the interaction 
can be described in the following way :
\begin{eqnarray}
 \label{op1}
 o_i(t+1) = \lambda [o_i(t)+\epsilon o_j(t)], \nonumber \\
 o_j(t+1) = \lambda [o_j(t)+\epsilon^{\prime} o_i(t)],
\end{eqnarray}
where $\epsilon$ and $\epsilon^{\prime}$ are uncorrelated random numbers between $0$ and $1$.

This type of interactions lead to a polarity or consensus formation depending upon the value of $\lambda$. 
The steady state average opinion after a long time $t$ would be given by 
$O = \sum_i |o_i|/N$. This represents the ``ordering'' in the system.
The system starts from a random disordered state ($O \sim 0$) and after a certain relaxation time $t = \tau$ 
moves to the ``para'' or 
``absorbing'' state where all individual agents have zero opinion
for $\lambda \leq 2/3$ or continuously changes to a ``symmetry broken'' 
or ``active'' state where all individuals
have opinion of same sign for $\lambda \geq 2/3$. The variance of $O$ shows a cusp near $\lambda = 2/3$.
The growth behaviour of the fraction of agents $p$ having extreme opinions
$o_i = \pm 1$ was found to be similar to $O$ \cite{soumya_1}. 
The relaxation time behaviour of the system shows a critical divergence of $\tau$, $\tau \sim |\lambda - \lambda_c|^{-z}$  for 
both $O$ and $p$ at 
$\lambda = \lambda_c = 2/3$. Values of $z$ for $O$ and $p$ are $1.0 \pm 0.1$ and $0.7 \pm 0.1$ 
respectively.

Notably, this model with interactions has a behaviour very similar to the simple iterative map,
\begin{equation}
 \label{map}
 y(t+1) = \lambda (1 + \epsilon_t ) y(t),
\end{equation}
with $y \leq 1$, where it was assumed that if $y(t) \geq 1$, $y(t)$ will be set equal to $1$. $\epsilon_t \in [0, 1]$ 
is a stochastic 
variable. In a mean-field approach Eq. \ref{map} reduces to
\begin{equation}
 y(t+1) = \lambda (1 + \langle \epsilon_t \rangle) y(t),
\end{equation}
where $\langle \epsilon_t \rangle = 1/2$. For $\lambda \leq 2/3$ $y(t)$ converges to $0$. 
An analytical derivation for the critical point was also given where it was found
that $\lambda_c = \exp \{-(2 \ln 2 - 1)\} \approx 0.6796$.

\subsubsection{Generalized LCCC model}
In the generalized LCCC model \cite{PS}, another parameter $\mu$ is introduced which is 
called the \textquoteleft influence\textquoteright parameter. 
It is a measure
of the influencing power or the ability of an individual to impose its opinion on
some other individual. 
Thus the interactions are described as folloows,
\begin{eqnarray}
 \label{op2}
 o_i(t+1) = \lambda_i o_i(t)+\epsilon \mu_j o_j(t),\nonumber \\
 o_j(t+1) = \lambda_j o_j(t)+\epsilon^{\prime} \mu_i o_i(t).
\end{eqnarray}
Note that here conviction and influence parameters of individual agents are different which lead to 
inhomogeneity in the society. In a simpler version, we may consider a homogeneous society so that all 
$\lambda$'s of different people are same. Also $\mu$'s for different people are same.

In this generalized version, the average opinion shows spontaneous symmetry breaking 
in the $\lambda-\mu$ plane. In the steady state the condition for non-zero solution of $O$ is 
\begin{equation}
(1-\lambda)^2 = \langle \epsilon \epsilon^{\prime} \rangle \mu^2 ,
\end{equation}
which gives that
``active'' and ``absorbing'' phases, separated by a phase boundary given by $\lambda = 1 - \mu/2$. 

\subsubsection{Other variants of the LCCC model}
Biswas et al. \cite{soumya_1} studied
some variants of the models discussed above. In one version, it was considered that when an individual $i$ meets  another individual $j$, she retains her own opinion proportional to her conviction parameter 
and picks up a random fraction of $j$'s opinion. Thus the interaction in equation form would now be,
\begin{eqnarray}
\label{op3}
 o_i(t+1) = \lambda o_i(t) + \epsilon o_j(t),\nonumber \\
 o_j(t+1) = \lambda o_j(t) + \epsilon^{\prime} o_i(t).
\end{eqnarray}

For $\lambda < \lambda_c$, for all agents $o_i = 0$ giving $O = 0$. For $\lambda > \lambda_c$, $O > 0$ and $O \rightarrow 1$ 
as $\lambda \rightarrow 1$. Numerical value of $\lambda_c = 1/2$. 
Mean field estimate gives for the stable value of $O$
\begin{equation}
 O(1-\lambda-\langle \epsilon \rangle)=0.
\end{equation}
Thus $\lambda_c = 1/2$.

Another variant of the LCCC model was studied \cite{soumya_1} with a slight modification to the original model that 
here a person in addition to being influenced by the interacting person's opinion, was influenced by the average 
opinion of the community. Thus the interaction equations read,
\begin{eqnarray}
\label{op4}
 o_i(t+1) = \lambda [o_i(t)+\epsilon o_j(t)] + \epsilon^{\prime} O(t), \nonumber \\
 o_j(t+1) = \lambda [o_j(t)+\eta o_i(t)] + \eta^{\prime} O(t).
\end{eqnarray}
The symmetric phase occurs for $\lambda \leq 1/3$ and symmetry-broken phase is obtained for $\lambda > 1/3$.

By a mean-field approach as $O$ reached a steady
state value,
\begin{equation}
O =\lambda(1+ \langle \epsilon \rangle) O + \langle \epsilon^{\prime} \rangle O,
\end{equation}
we have $\lambda_c = 1/3$.
In all these models, the
critical exponents associated with the physics of phase transitions were all estimated. 

\subsubsection{Discrete LCCC model}
In the discrete version of LCCC model one considers that opinions can take only discrete values, i.e., $o_i$ can 
take only three values $[o_i \in \{-1,0,+1\}]$. This particular 
version of the LCCC model was exactly solved \cite{soumya_2}, which also showed
an ``active-absorbing phase transition'' as was seen in the continuous version\cite{Lalla1,Lalla2}. Apart
from the two-agent or binary interaction, the three-agent interaction were also taken
into account. While the phase diagram of the two-agent interaction led to a continuous
transition line, the three-agent interaction showed a discontinuous transition \cite{soumya_2}.

\subsubsection{Disorder Induced Phase Transition in Kinetic Exchange Models of Opinion Formation}
In this model of continuous opinion dynamics, both positive and negative mutual interactions
were studied \cite{soumya_3}. The interaction equations are as follows :
\begin{eqnarray}
 \label{op5}
 o_i(t+1) = o_i(t) + \mu_{ij} o_j(t),\nonumber \\
 o_j(t+1) = o_j(t) + \mu_{ij} o_i(t),
\end{eqnarray}
where $\mu_{ij}$ are randomly chosen to be either $+1$ or $-1$.
Negative interactions are included here with probability
$p$, the role of which is like a disordering field, similar to temperature in thermally driven phase transitions.
Beyond a particular value $p=p_c$ a phase transition from an ordered phase to a disordered phase occurs.
Results from exact
calculation of a discrete version also shows the phase transition at $p_c$.
\subsubsection{LCCC model with bounded confidence}
\begin{figure}
 \includegraphics[scale=0.45, angle=-90]{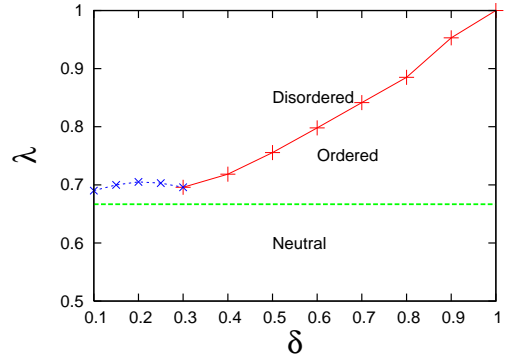}
 \caption{Phase diagram on the $\delta-\lambda$ plane. Plot shows the existence of the neutral region for
$\lambda \leq \lambda_{c1} \simeq 2/3$, the ordered region and the disordered region. The ordered
and disordered regions are separated by a first order boundary for $\delta \geq 0.3$. For
$\delta < 0.3$, the phase boundary has been obtained approximately only from the behaviour of the order parameter.
Taken from \cite{PS2}.
}
\label{delta_lambda_plane}
\end{figure}

In this restricted LCCC model  \cite{PS2}, two agents
interact according to Eq. \ref{op1} only when $|o_i-o_j| \leq 2\delta$, where the parameter $\delta \in [0, 1]$  
represents the \textquoteleft confidence\textquoteright level. There are
two extreme limits corresponding to this model: 
\begin{enumerate}
 \item $\delta = 1$ which brings it back to the original model.
LCCC model
\item $\delta = 0$ which is the case when two agents interact only when their
opinions are exactly same.
\end{enumerate}
Three different states were defined to identify the status
of the system. 
\begin{itemize}
 \item Neutral State : When $o_i = 0$ for all $i$, the state is called neutral state.
 \item Disordered State : $o_i = 0$ for all $i$, but $O \sim 0$, the state is called disordered state.
 \item Ordered State : when $O = 0$ corresponding state is called an ordered state.
\end{itemize}
The three states are located in the $\delta-\lambda$ plane. The ordered and disordered regions in the plane
are separated by a first order boundary (continuous line in red) for $\delta \geq 0.3$ (obtained using a finite
size scaling analysis). For $\delta < 0.3$, the phase boundary (broken line in blue) has been obtained
approximately only from the behaviour of the order parameter (Fig. \ref{delta_lambda_plane}).

\subsubsection{Percolation in LCCC model}
The opinion spreading among people in a society may be compared
to the percolation problem in physics. The agents
are assumed to be placed on the sites of a square lattice and follow the
LCCC dynamics \cite{chandra}. A geometrical cluster consisting of the adjacent sites having opinion value more
than or equal to a predefined threshold value $\Omega$. At steady state, the percolation
order parameter is measured. At a particular value of $\lambda = \lambda_c^p$, the system undergoes a percolation transition.
As $\Omega$ decreases, $\lambda_c^p$ also and approaches $\lambda_c$ as $\Omega \rightarrow 0$. The critical 
exponents are independent of $\Omega$ as well as $\lambda$ and
$\mu$. The critical exponents are significantly different from those obtained for
static and dynamic Ising system and standard percolation. The exponents suggest that this LCCC
model has a separate universality class from the viewpoint of percolation transition.

\subsubsection{Damage spreading in Model of Opinion Dynamics}
The damage spreading phenomena was studied in the opinion dynamics model proposed in \cite{soumya_3} 
in two ways, 
\begin{itemize}
 \item Traditional Method (TM) : In this method, two systems of $N$ individuals are simulated 
 using the same initial random 
opinions either discrete or continuous, except for one randomly chosen individual. The two systems are then allowed to
evolve using same random numbers. 
 \item Nature versus Nurture Method (NVN) : In this (NVN) method, the initial systems are identical but different 
 random numbers are used for the time evolution.
\end{itemize}
In both cases, a damage spreading transition occurs at $p_d$ where $p_d \approx 0.18$ for TM
and $p_d = 0$ for NVN \cite{khaleque}. Here it is found that $p_d < p_c$, the order-disorder trantion point.
The result signifies that for TM, for $p_d < p < p_c$ , even
when consensus is reached, if we make very small changes even in 
a single agent, there is always a finite probability that the system leads 
to a different consensus state. In NVN, $p_d = 0$ signifies that
if the same agent goes through a different sequence of
interactions, the result will be different for any $p$ with
finite probability. However, the dynamics of the damage shows a non-monotonicity
making it difficult to comment on the exact nature of damage or to estimate the exponents related to it.

\section{Summary and discussions}
We briefly described here, the kinetic exchange models for economics and sociology and some applications 
derived from these models. Taking inspiration from kinetic theory of gas molecules, a purely statistical system,  
these kind of models give an idea of how completely different systems might
lead to similar or emergent collective behaviour, as they have some similar connections in the microscopic units. 
However, due to such ``micro-oriented'' framework one overlooks the system-wide effects which
can be very important for a real economy and society. However, one should bear in mind that whatever we discussed here in this paper, is to a large extent idealistic. A real economy is much more complex than any or all of these
models. In case of a real economy, minute changes in the characteristics of the agents or firms, or simply the addition or deletion of a link of the socio-economic network, can alter the emergent behaviour to a great extent. Models originating from simple multi-agent models such as the ones described here, should be extended to incorporate such features and emergent behaviours, which might help one to understand many real-life economic phenomena or even the financial crisis, such as the one observed during $2007-2008$.

It should also be borne in mind that besides being models of idealized economy or society, these simple models have a very nice mathematical or statistical appeal. Mathematicians, physicists, and economists, have tried to play around with these models (or their variants) and studied the associated non-linear dynamics, steady-state behaviours and related questions. Apenko \cite{apenko} used a different approach and proved the monotonic entropy growth for a nonlinear discrete-time model of a random market, based on binary collisions, which may be also viewed as a particular case of the Ulam's redistribution of energy problem. 
In that study, a single step of the nonlinear evolution was treated as a combination of two steps, 
first one is related to an auxiliary linear two-particle process and second one is a kind of a coarse-graining. 
It was shown that on both steps the entropy increases. Therefore he concluded that the entropy is 
indeed monotonically increasing for the original nonlinear
problem. A similar entropy approach was followed in \cite{ruiz}, where they considered
different versions of a continuous economic model, which
takes into account some idealistic characteristics of the markets and agents randomly exchange in pairs, and their functional mappings. They showed that 
the system had a fixed point which can be reached asymptotically following a trajectory of monotonically 
increasing entropy which takes its maximum
value on the equilibrium. In this manner, 
the existence of an H-theorem could be computationally checked.

\begin{acknowledgement}
The authors would like to thank all their collaborators and students, whose works
have been presented here. 
\end{acknowledgement}
%

%


\begin{thebibliography}{99.}
%
\bibitem{Mandl} F.~Mandl, \textit{Statistical Physics} (2nd Ed.) (John Wiley, New York, 2002).

\bibitem{terHaar1995} D. ter Haar, {\em Elements of Statistical Mechanics} (Butterworth-Heinemann, Oxford,1995).

\bibitem{Sethna_A} J.P. Sethna, {\em Statistical Mechanics} (Oxford University Press, Oxford, 2006).

\bibitem{SCCC}
S.~Sinha, A.~Chatterjee, A.~Chakraborti and B.K.~Chakrabarti, 
\textit{Econophysics: An Introduction} (Wiley-VCH, Berlin, 2010).

\bibitem{sen_chak_book}P.~Sen and B.K.~Chakrabarti, \textit{Sociophysics : An Introduction} (Oxford University Press, Oxford, 2013).

\bibitem{ball} P.~Ball, Physica A \textbf{314}, 1 (2002).

\bibitem{CUP}B.K. Chakrabarti, A. Chakraborti, S.R. Chakravarty and A. Chatterjee, \textit{Econophysics of Income and Wealth Distributions} (Cambridge University Press, Cambridge, 2013).

\bibitem{Pareto:1897}V.~Pareto, \textit{Cours d'economie Politique} (F. Rouge, Lausanne, 1897).

\bibitem{Souma} W. Souma, Fractals \textbf{9}, 463 (2001).

\bibitem{Matteo} T.~Di Matteo, T.~Aste and S.~T.~Hyde, 
in Eds. F.~Mallamace and H.~E.~Stanley, \textit{The Physics of Complex Systems (New Advances and Perspectives)} (IOS Press, Amsterdam, 2004), p.~435.

\bibitem{Clementi} F. Clementi and M. Gallegati, Physica A \textbf{350}, 427 (2005).

\bibitem{Sinha} S. Sinha, Physica A \textbf{359}, 555 (2006).

\bibitem{Silva} A.C.~Silva and V.M.~Yakovenko, Europhys. Letts. \textbf{69}, 304 (2005).

\bibitem{Dragu}A.A.~Dr\u{a}gulescu and V.M.~Yakovenko, Physica A \textbf{299}, 213 (2001).

\bibitem{Mandelbrot:1960}B.B.~Mandelbrot, Int. Econ. Rev. \textbf{1}, 79 (1960).

\bibitem{EWD05} Eds. A.~Chatterjee, S.~Yarlagadda and B.~K.~Chakrabarti
\textit{Econophysics of Wealth Distributions} (Springer Verlag, Milan, 2005).

\bibitem{ESTP} Eds. B.K.~Chakrabarti, A.~Chakraborti and A.~Chatterjee,
\textit{Econophysics and Sociophysics} (Wiley-VCH, Berlin, 2006).

\bibitem{Yakovenko:RMP}
V.M.~Yakovenko and J.~Barkley~Rosser,~Jr., Rev. Mod. Phys. \textbf{81},  1703 (2009).

\bibitem{datapap}
A.A.~Dr\u{a}gulescu and V.M.~Yakovenko, Eur. Phys. J. B \textbf{20}, 585 (2001);
M.~Levy and S.~Solomon, Physica A \textbf{242}, 90 (1997);
H.~Aoyama, W.~Souma and Y.~Fujiwara, Physica A \textbf{324}, 352 (2003);
N.~Ding and Y.~Wang, Chinese Phys. Letts. \textbf{24}, 2434 (2007).

\bibitem{marjitIspolatov}
B.K.~Chakrabarti and S.~Marjit, Ind. J. Phys. B \textbf{69}, 681 (1995);
S.~Ispolatov, P.L.~Krapivsky and S.~Redner, Eur. Phys. J. B \textbf{2}, 267 (1998).

\bibitem{Dragulescu:2000}
A.A.~Dr\u{a}gulescu and V.M.~Yakovenko, Eur. Phys. J. B \textbf{17}, 723 (2000).

\bibitem{Chakraborti:2000}
A.~Chakraborti and B.K.~Chakrabarti, Eur. Phys. J. B \textbf{17}, 167 (2000).

\bibitem{Chatterjee:rev}
A.~Chatterjee and B.K.~Chakrabarti, Eur. Phys. J. B \textbf{60}, 135 (2007);
A.~Chatterjee, S.~Sinha and B.K.~Chakrabarti,
Curr. Sci. \textbf{92}, 1383 (2007).

\bibitem{Chakrabarti:2010}
A.S.~Chakrabarti and B.K.~Chakrabarti, Economics E-journal \textbf{4} (2010); available at  http://www.economics-ejournal.org/economics/journalarticles/2010-4.

\bibitem{Chatterjee:2010}
A.~Chatterjee, in Eds. G.~Naldi et. al., \textit{Mathematical Modeling of Collective Behavior in 
Socio-Economic and Life Sciences} (Birkha\"{u}ser, Boston, 2010), ~p. 31. 

\bibitem{Chakraborti2002} A.~Chakraborti, Int. J. Mod. Phys. C \textbf{13}, 1315 (2002).

\bibitem{Hayes} B. Hayes, American Scientist \textbf{90}, 400 (2002).

\bibitem{lopez} R.~L\'{o}pez-Ruiz, J.L.~L\'{o}pez and X.~Calbet,
ESAIM Proceedings \textbf{36}, 189 (2012).

\bibitem{Patriarca:2004}
M.~Patriarca, A.~Chakraborti and K.~Kaski, Phys. Rev. E \textbf{70}, 016104 (2004).

\bibitem{Repetowicz:2005}
P.~Repetowicz, S.~Hutzler and P.~Richmond, Physica A \textbf{356}, 641 (2005).

\bibitem{Lallouache:2010}
M.~Lallouache, A.~Jedidi and A.~Chakraborti, Science and Culture \textbf{76}, 478 (2010).

\bibitem{Angle}
J.~Angle, Social Forces \textbf{65}, 293 (1986); Physica A \textbf{367}, 388 (2006).

\bibitem{ChakrabartiABK} A.~Chakraborti and M.~Patriarca, Pramana J. Phys. \textbf{71}, 233 (2008).

\bibitem{Chakraborti-prl} A. Chakraborti and M. Patriarca, Phys. Rev. Lett. \textbf{103}, 228701 (2009).

\bibitem{AJP} M. Patriarca and A. Chakraborti, Am. J. Phys. \textbf{81}, 618 (2013).

\bibitem{Chatterjee:2004}
A.~Chatterjee, B.K.~Chakrabarti and S.S.~Manna, Physica A \textbf{335}, 155 (2004);
Phys. Scr. T \textbf{106}, 36 (2003).

\bibitem{Chatterjee:2005}
A.~Chatterjee, B.K.~Chakrabarti and R.B.~Stinchcombe, Phys. Rev. E \textbf{72}, 026126 (2005).

\bibitem{Mohanty:2006}
P.K.~Mohanty, Phys. Rev. E \textbf{74}, 011117 (2006).

\bibitem{Kargupta}
A.~Kar~Gupta, in Ref.~\cite{ESTP}, p.~161.

\bibitem{ecoanneal}
A.~Chatterjee and B.K.~Chakrabarti, Physica A \textbf{382}, 36 (2007).

\bibitem{Toscani}
B. D\"{u}ring and G. Toscani, Physica A \textbf{384}, 493 (2007);
B. D\"{u}ring, D. Matthes and G. Toscani, Phys. Rev. E \textbf{78}, 056103 (2008);
D. Matthes and G. Toscani, J. Stat. Phys. \textbf{130}, 1087 (2008);
D. Matthes and G. Toscani, Kinetic and related Models \textbf{1}, 1 (2008);
V. Comincioli, L. Della Croce and G. Toscani, Kinetic and Related Models \textbf{2}, 135 (2009).

\bibitem{Chatterjee:2009}
A.~Chatterjee, Eur. Phys. J. B \textbf{67}, 593 (2009).

\bibitem{ChakrabartiASBK}
A.S.~Chakrabarti and B.K.~Chakrabarti, Physica A \textbf{388}, 4151 (2009).

\bibitem{ChakrabartiASBK2} A.S.~Chakrabarti and B.K.~Chakrabarti, Physica A \textbf{389}, 3572 (2010).

\bibitem{manna} A.~Chakraborty and S.S.~Manna, Phys. Rev. E \textbf{81}, 016111 (2010).

\bibitem{iglesias} J.R. Iglesias, Science and Culture (Kolkata) \textbf{76} , 437
(2010); S. Pianegonda, J.R. Iglesias, G. Abramson and J.L. Vega, Physica A \textbf{322}, 667 (2003).

\bibitem{AGhosh} A.~Ghosh (unpublished); A. Ghosh, A.S. Chakrabarti, A.K. Chandra and A. Chakraborti, in Eds. F. Abergel, H. Aoyama, B.K. Chakrabarti, A. Chakraborti and A. Ghosh, \textit{Econophysics of Agent-Based Models} (Springer, Milan, 2014), pp 99-129.

\bibitem{sinha_chat} Independent studies (unpublished) made by S. Sinha and also A. Chatterjee in 2005-2006, had similar observations; private communications.

\bibitem{threshold}A.~Ghosh, U.~Basu, A.~Chakraborti and B.K.~Chakrabarti, Phys. Rev. E \textbf{83}, 061130 (2011).

\bibitem{Chatt-sen:2010}
A.~Chatterjee and P.~Sen, Phys. Rev. E \textbf {82}, 056117 (2010).

\bibitem{goswami} S. Goswami, P. Sen and A. Das, Phys. Rev. E \textbf{81}, 021121 (2010).


\bibitem{pkm} P.K.~Mohanty, Phys. Rev. E \textbf{74}, 011117 (2006).

\bibitem{chak_int2} A.S.~Chakrabarti, Eur. Phys. J. B \textbf{86}, 255 (2013); A.S.~Chakrabarti, Physica A \textbf{391}, 6039 (2012).

\bibitem{Castellano} C.~Castellano, S.~Fortunato and V.~Loreto, Rev. Mod. Phys. \textbf{81}, 591 (2009).

\bibitem{Galam} S.~Galam, \textit{Sociophysics: A Physicist's Modeling of Psycho-Political Phenomena (Understanding Complex Systems)} (Springer, Heidelberg, 2012).

\bibitem{Stauffer} D.~Stauffer, J. Stat. Phys. \textbf{151}, 9 (2013).

\bibitem{Lalla1} M.~Lallouache, A.~Chakraborti and B.K.~Chakrabarti, Science and Culture \textbf{76}, 485 (2010).

\bibitem{Lalla2} M.~Lallouache, A.S.~Chakrabarti, A.~Chakraborti and B.K.~Chakrabarti, Phys. Rev. E \textbf{82}, 056112 (2010).

\bibitem{soumya_1} S.~Biswas, A.K.~Chandra, A.~Chatterjee and B.K.~Chakrabarti, J. Phys. Conf. Ser. \textbf{297}, 012004 (2011).

\bibitem{PS} P.~Sen, Phys. Rev. E \textbf{83}, 016108 (2011).

\bibitem{soumya_2} S.~Biswas, Phys. Rev. E \textbf{84}, 056106 (2011).

\bibitem{soumya_3} S.~Biswas, A.~Chatterjee and P.~Sen, Physica A \textbf{391}, 3257 (2012).

\bibitem{PS2} P.~Sen, Phys. Rev. E \textbf{84}, 016115 (2012).

\bibitem{chandra} A.K.~Chandra, Phys. Rev. E \textbf{85}, 021149 (2012).

\bibitem{khaleque} A.~Khaleque and P.~Sen, available at arXiv: 1312.7718v1 (2013).

\bibitem{apenko} S.~Apenko, Phys. Rev. E \textbf{87}, 024101 (2013).

\bibitem{ruiz} R.~L\'{o}pez-Ruiz, E.~Shivanian, J.L.~L\'{o}pez, available at arXiv:1307.2169 (2013).
 
\end{thebibliography}
\end{document}